\documentclass[%
 aip,
 amsmath,amssymb,
 reprint,%
]{revtex4-1}
\pdfoutput=1
\usepackage[latin9]{inputenc}
\usepackage[english]{babel}
\usepackage{graphicx}
\usepackage{bm}
\usepackage{color,soul}
\usepackage{dcolumn} 
\usepackage{amsfonts}
\usepackage{natbib}
\usepackage{upgreek}
\usepackage{amssymb}
\usepackage{amsmath}
\usepackage{physics}
\usepackage{graphicx}
\usepackage{lipsum}
\usepackage{float}
\usepackage[dvipsnames]{xcolor}
\usepackage{url}
\usepackage{hyperref}
\usepackage[normalem]{ulem}
\hypersetup{
    colorlinks=true,
    urlcolor=blue,
    linkcolor=blue,
    citecolor=blue}
\usepackage{upgreek}

\newcommand{\otherlabel}[2]{\protected@edef\@currentlabel{#2}\label{#1}}

\begin{document}

\title{Lattice structure dependence of laser-induced ultrafast magnetization switching in ferrimagnets}


\author{J. A. V\'elez}
\affiliation{Donostia International Physics Center, 20018 San Sebasti\'an, Spain}
\affiliation{Polymers and Advanced Materials Department: Physics, Chemistry, and Technology, University of the Basque Country, UPV/EHU, 20018 San Sebasti\'an, Spain}
\author{R. M. Otxoa}
\affiliation{Hitachi Cambridge Laboratory, J. J. Thomson Avenue, Cambridge CB3 0HE, United Kingdom\looseness=-1}
\affiliation{Donostia International Physics Center, 20018 San Sebasti\'an, Spain}
\author{U. Atxitia}
\email{u.atxitia@csic.es}
\affiliation{Instituto de Ciencia de Materiales de Madrid, CSIC, Cantoblanco, 28049 Madrid, Spain\looseness=-1}

\date{\today}

\begin{abstract}
The experimental discovery of single-pulse ultrafast magnetization switching in ferrimagnetic alloys, such as GdFeCo and MnRuGa, opened the door to a promising route toward faster and more energy efficient data storage. 
A recent semi-phenomenological theory has proposed that a fast, laser-induced demagnetization below a threshold value puts the system into a dynamical regime where angular momentum transfer between sublattices dominates.  
Notably, this threshold scales inversely proportional to the number of exchange-coupled nearest neighbours considered in the model, which in the simplest case is directly linked to the underlying lattice structure.   
In this work, we study the role of the lattice structure on the laser-induced ultrafast magnetization switching in ferrimagnets by complementing the phenomenological theory with atomistic spin dynamics computer simulations.  
We consider a spin model of the ferrimagnetic GdFeCo alloy with increasing number of exchange-coupled neighbours. 
Within this model, we demonstrate that the laser-induced magnetization dynamics and switching depends on the lattice structure. Further, we determine that the critical laser energy for switching reduces for decreasing number of exchange-coupled  neighbours.  
\end{abstract}

\maketitle

Fast, reliable, and inexpensive data manipulation and storage is the cornerstone for innovation and progress of our information-technology-based society. Ultrafast magnetism holds promise for fast and low energy data manipulation solutions\cite{KirilyukRMP2010,Carva2017,El-Ghazaly2020,barker2021,DaviesJMM2022}. The field was initiated by the discovery of femtosecond laser pulse induced subpicosecond demagnetization in Ni\cite{Beaurepaire1996}. To this breakthrough followed the demonstration of field-free magnetization switching in GdFeCo alloys using a train of circularly polarized pulses\cite{StanciuPRL2007}. Later on, a combined theoretical/experimental study showed the possibility of single pulse switching using  linearly polarized light\cite{RaduNature2011,OstlerNatComm2012}. This finding uncovered the purely thermal origin of the switching process, which in turn was  used to demonstrate that ultrafast heating by picosecond electric pulses is sufficient to achieve switching in GdFeCo \cite{Yang2017}. Single-pulse switching can also be accomplished in CoTb alloys \cite{LiuAPL2023}, Gd-based ferrimagnetic multilayers\cite{Lalieu2017,Beens2019}, magnetic tunnel junctions of Tb/Co \cite{AvilesSREP2020}, and the rare-earth-free Heusler alloy Mn$_2$Ru$_x$Ga \cite{Banerjee2020,Jakobs2022}. For future information technologies ultrafast magnetization manipulation promises high potential, as such, further understanding of the microscopic origin of single-pulse magnetization switching is key  for ultrafast spintronics applications \cite{Lalieu2019}. 

Single pulse magnetization switching in ferrimagnets has been described using computer simulations based on atomistic spin dynamics (ASD)\cite{WienholdtPRB2013,Chimata2015,Jakobs2021,Ceballos2021} and phenomenological models \cite{Mentink2012,Atxitia2016,Mentink2017,GRIDNEV2021}.
A recent work has merged these models into an unified macroscopic theory that describes magnetization dynamics and switching of two-sublattice ferrimagnets upon femtosecond laser excitation\cite{JakobsPRB2022,JakobsPRL2022}. Within this theory, the switching process becomes possible due to an enhancement of the exchange relaxation -- angular momentum exchange between sublattices -- when the magnetization of the sublattices is reduced below a certain threshold that depends on material parameters. 
After femtosecond laser photo-excitation, the electron system enters a high temperature regime at which angular momentum dissipation into electron or phonon degrees of freedom dominates. This so-called relativistic relaxation is related to the spin-orbit coupling connecting spin and orbital degrees of freedom. Depending on the laser power, the  sublattice magnetization can reduce down the threshold that leads to magnetic switching.   
Once the laser pulse is gone, the electron system starts to cool down, which leads to a local recovery of magnetic order  due to the exchange coupling between spins. 
In two sublattice magnets, the so-called exchange relaxation, through local exchange of angular momentum between sublattices, drives magnetic switching.  
Previous computational works using ASD methods have investigated how switching depends on a variety of parameters, such as element-specific damping \cite{Jakobs2021,Ceballos2021}, rare-earth concentration\cite{LiuAPL2023}, duration of the laser pulse \cite{Davies2020,Jakobs2021} or the role of the initial temperature\cite{Barker2013}. The impact of the number of exchange-coupled neighbours on the switching behaviours in GdFeCo has remained however unexplored. 

In this work, we provide insights about how the single-pulse magnetic switching of the ferrimagnetic alloy GdFeCo depends on the lattice structure, in terms of number of exchange-coupled spins. We use both a semi-phenomenological theory as well as atomistic  spin dynamics simulations to demonstrate that by reducing the number of exchange-coupled spins, the laser energy necessary to switch the magnetic state of GdFeCo reduces significantly.  

The magnetization dynamics of a ferrimagnet composed of two magnetic sublattices, such as GdFeCo, can be described in terms of the sublattice angular momentum $S_a = \mu_a \langle s_a \rangle /\gamma$, where $m_a=\langle s_a \rangle$ and $\mu_a$ its atomic magnetic moment\cite{MentinkPRL2012,JakobsPRL2022}
\begin{equation}
\frac{d S_a}{d t} =  \alpha_{a} \mu_a  H_a + \alpha_{\rm{ex}} ( \mu_a H_a - \mu_b H_b )
\label{eq:long-LLB-1-ex}
\end{equation}
where $a=$ Fe and $b=$ Gd. Here, $\alpha_a$ stands for the macroscopic relativistic damping parameter\cite{JakobsPRB2022}
\begin{equation}
\alpha_{a} = 2\lambda_a \frac{L(\xi_a)}{\xi_a}.
\label{eq:alpha-a}
\end{equation}
Here, $L(\xi)$ stands for the Langevin function and $\lambda_a$ is the element-specific intrinsic damping parameter. The so-called thermal field is defined as $\xi_a= \beta \mu_a H_a^{\rm{MFA}}$ ($\beta=1/k_{\rm{B}} T$) within the mean-field approximation (MFA), where   
\begin{equation}
\mu_a H_a^{\rm{MFA}} = z_a J_{aa} m_a + z_{ab} J_{ab} m_b.
\label{eq:MFA-LLB}
\end{equation}
The non-equilibrium fields are defined as 
\begin{equation}
H_a = \frac{(m_a-m_{0,a})}{\mu
_a\beta L'(\xi_a)},
\label{eq:muaHa}
\end{equation}
where, $L'(\xi) = dL/d\xi$ and $m_{0,a}=L(\xi_a)$. We note that in the underlying model behind the MFA the Gd spins are randomly located in a regular Fe spin lattice with a concentration $q$, and $z$ corresponds to the number of exchange-coupled neighbours. Thus, $z_a=z q$ represents the average number of nearest neighbours (n.n.) of spins of type $a$, and similarly $z_{ab}=z(1-q)$ represents the average number of n.n. of type $b$.
Since single-pulse switching has been observed mostly for Gd concentration of around $25\%$, in our model we restrict to that concentration of Gd spins. 
Within this model, one can define $J_{0,a}=z qJ_{aa}$ and $J_{0,ab}=z(1-q)J_{ab}$, that is it, $\mu_a H_a^{\rm{MFA}}=J_{0,a}m_a+J_{0,ab}m_b$. Within the MFA the equilibrium magnetization $m_e=L(\xi_e)$ only depends on the values of $J_{0,a}$ and $J_{0,ab}$ but it is insensitive to the lattice structure. In comparison to the MFA, ASD simulations can capture the small differences coming from the lattice structure and as a consequence the equilibrium magnetization slightly depends on the lattice structure \cite{GaraninPRB1996}. For the MFA calculations we assume a fixed values for $J_{0,a}$ and $J_{0,ab}$ for all values of $z$. These assumptions give the same temperature-dependent equilibrium magnetization for all cases. 

The number of exchange-coupled neighbours $z$ shows up explicitly in the so-called exchange relaxation parameter
\begin{equation}
\alpha_{\rm{ex}} = \frac{1}{2z}\left(\frac{\alpha_{a}}{m_a} +  \frac{\alpha_{b}}{m_b} \right).
\label{eq:alpha-ex}
\end{equation}
The number of neighbours $z$ will therefore become relevant for the description of the magnetization dynamics. For example, in the exact MFA limit, where $z\rightarrow{\infty}$ while $J_{0,a}$ remains constant, $\alpha_{\rm{ex}}\rightarrow{0}$. 
The vanishing  of the exchange relaxation parameter for large number of exchange-coupled neighbours directly affects the ability of the system to switch. By contrast, the exchange relaxation parameter increases as the number of neighbours reduce, which could make the switching process more energy efficient.  

As a first approximation, we can assume that the magnetization relaxation pathway, either of relativistic or exchange nature, is determined by the value of their corresponding relaxation  parameters. Therefore, the crossover between relativistic- to exchange-dominated regimes can be estimated by finding the conditions at which $\alpha_{\rm{ex}}>\alpha_a$.  For simplicity we consider first the case $\lambda_a=\lambda_b$ in Eq. \eqref{eq:alpha-a}, for which $\alpha_a \approx \alpha_b=\alpha$\cite{Atxitia2012}. Under this assumption, one can simplify the condition $\alpha=\alpha_{\rm{ex}}$ to 
\begin{equation}
    z= \frac{1}{m_a} +\frac{1}{m_b}
\end{equation}
\begin{figure}[!t]
\includegraphics[width=4.2cm]{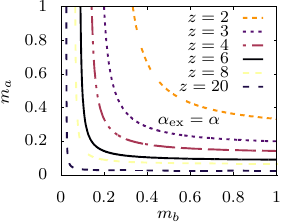}
\includegraphics[width=4.2cm]{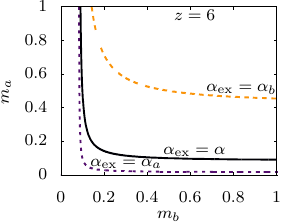}
\caption{(a) Lines separate the regions where $\alpha_{\rm{ex}}<\alpha$ (right side) and the region where $\alpha_{\rm{ex}}>\alpha$ (left side) for a range of nearest neighbours number $z$ ($\lambda_a = \lambda_b$). (b) Similar for the case $z=6$, for $\alpha_a = \alpha_b$ coincident to (a), and two lines corresponding to $\alpha_a/\alpha_b = 5$, one for the case $\alpha_{\rm{ex}}=\alpha_a$ and another for $\alpha_{\rm{ex}}=\alpha_b$.}
\label{im:0}
\end{figure} 
Figure \ref{im:0} (a) shows the lines separating the regions $\alpha_{\rm{ex}}<\alpha$ and $\alpha_{\rm{ex}}>\alpha$. As the number of neighbours $z$ reduce, the region $(m_a,m_b)$ for which $\alpha_{\rm{ex}}>\alpha_a$ increases. For instance, for the limit case of $z=2$ (spin chain), $\alpha_{\rm{ex}}=\alpha$ already for relatively large values of $m_a$ and $m_b$. For a larger number of neighbours $z$, corresponding to simple cubic ($z=6$) or face-centered cubic ($z=12$) lattices, $\alpha_{\rm{ex}}$ is relatively smaller than $\alpha$ for a large region $(m_a,m_b)$.

Experimental observations combined with ASD simulations suggest that in rare earth transition metal alloys, the damping values are element specific\cite{Ceballos2021,Jakobs2021}. Specifically, it was found that using $\lambda_{\rm{Fe}}=0.06$ and $\lambda_{\rm{Gd}}=0.01$ in the ASD simulations, one can qualitatively reproduce the ultrafast magnetization dynamics and switching of the GdFeCo alloys in a range of Gd concentrations\cite{Jakobs2021}.  Similarly, in Gd$_{22-x}$Tb$_x$Co$_{78}$ alloys, it was found element-specific damping values  ($\lambda_{\rm{Co}}=\lambda_{\rm{Tb}}=0.05$ and $\lambda_{\rm{Gd}}=0.005-0.05$) could describe switching as a function of Tb content\cite{Ceballos2021}. For the sake of simplicity, we illustrate the impact of element-specific damping on the relation between relaxation parameters for a particular case, $z=6$. Figure \ref{im:0} (b) shows $\alpha_{\rm{ex}}=\alpha$ in solid lines, which corresponds to Fig. \ref{im:0}(a), and $\alpha_a /\alpha_b = 5$, which agree to experimental observations. For element-specific damping values, each element (sublattice) will enter the exchange dominated regime under different conditions, namely, sublattice $a$ when  $\alpha_{\rm{ex}}=\alpha_a$ and sublattice $b$ when $\alpha_{\rm{ex}}=\alpha_b$. Our model predicts that under those circumstances the magnetization relaxation of the Gd sublattice is dominated by the exchange relaxation during the whole demagnetization process, i.e.,  by transfer of angular momentum to the Fe sublattice, whereas Fe sublattice angular momentum relaxation remains mostly relativistic -- transfer of angular momentum to other degrees of freedom.

\begin{figure}[!t]
\includegraphics[width=8.6cm]{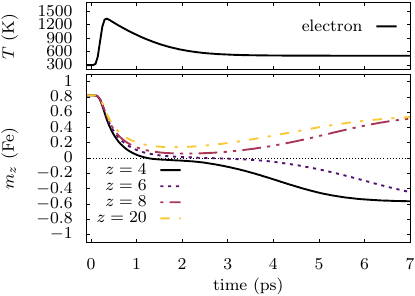}
\caption{(Top) Electron temperature dynamics driven by a femtosecond laser pulse. (Bottom) Iron sublattice magnetization dynamics and switching for systems with different number of exchange-coupled nearest neighbours $z$. The sublattice magnetization dynamics is calculated using Eq. \eqref{eq:long-LLB-1-ex} with the electron temperature profile in the top figure as an input.}
\label{im:1}
\end{figure} 
It is worth noting that the magnetization relaxation dynamics described by Eq. \eqref{eq:long-LLB-1-ex} not only scales with the value of the relaxation parameters but also with the non-equilibrium effective field $H_a$ [Eq. \eqref{eq:muaHa}]. In particular, the relaxation of the sublattice magnetization Eq. \eqref{eq:long-LLB-1-ex} can be split into two contributions, the relativistic relaxation rate: $\Gamma^r_a =\alpha_a\mu_a H_a$, and exchange relaxation rate $\Gamma^{\rm{ex}}=\alpha_{\rm{ex}}(\mu_a H_a-\mu_b H_b)$. 
After the application of a laser pulse the temperature quickly increases beyond the critical temperature and the dynamics is dominated by the thermal fields, which translates into $\Gamma^r_a \sim \lambda_a k_{\rm{B}}Tm_a$ whereas $\Gamma^{\rm{ex}}_{a} \sim \lambda_a k_{\rm{B}} T/z$. Thus, a prerequisite for switching is that the laser pulse produces a high temperature profile (see Fig. \ref{im:1} (Top)) to quickly reduce the magnetization $m_a$, so that the exchange relaxation takes over, $\Gamma^{\rm{ex}} > \Gamma_a^r$. 
One would expect a more efficient switching process for lower values of $z$, and a highly non-efficient for high values of $z$.
One can rationalize this by analysing the equations of motion for some limiting situations.
For the same value of the intrinsic damping parameter $\lambda_{\rm{Fe}}=\lambda_{\rm{Gd}}$, by assuming that one of the sublattice demagnetizes faster (Fe in GdFe alloys) than the other, soon after the application of a fs laser pulse one finds that $m_a\ll m_b$, from the condition $\Gamma^{\rm{ex}} > \Gamma^r_{a}$, one gets  $m_a < \sqrt{m_b/2z}$, and from $\Gamma^{\rm{ex}} > \Gamma^r_{b}$ one gets $m_a \leq 1/2z$\cite{JakobsPRL2022}. 
For example, for a lattice with fcc+bcc structure, $z=20$, $m_a< \sqrt{1/20} = 0.15$, while for a two-dimensional magnet with a square lattice structure, $z=4$, $m_a< \sqrt{1/8} = 0.35$, or a spin chain, $z=2$, $m_a< \sqrt{1/4} = 0.5$. 

This finding is illustrated in Fig. \ref{im:1} (Bottom), where the magnetization dynamics of each sublattice  is calculated using Eq. \eqref{eq:long-LLB-1-ex} coupled to the so-called two-temperature model (TTM), which describes the dynamics of the electron and phonon temperature \cite{Kaganov1957,Chen2006}. For all $z$ values, we use the same parameters for the TTM, and thus the electron and phonon temperatures dynamics are the same (Fig. \ref{im:1} (Top)). In the TTM, we use the following values for the electron specific heat $c_e=\gamma_e T_e$ ($\gamma_e=700$ J/Km$^3$), phonon specific heat $c_{\rm{ph}}=3 \times 10^{6}$  J/Km$^3$ and electron-phonon coupling  $g_{\rm{e-ph}}=6 \times 10^{17}$  J/sKm$^3$ \cite{JakobsPRL2022}. The heat-bath to which the spins are coupled is represented by the electron system. Figure \ref{im:1}(Bottom) shows the magnetization dynamics of the Fe sublattice for different number of neighbours $z$. We set the exchange parameters in such a way the Curie temperature of all systems is the same. For the same energy input from the laser pulse, the system with less exchange-coupled spins $z=4$, switches faster than for example $z=6$.  For larger number of $z$, switching does not occur since the magnetization threshold for switching is not achieved. In the limit $z\rightarrow \infty$, the exchange relaxation parameter is null, and consequently switching becomes impossible.  

Within the MFA used so far, the temperature-dependence of the equilibrium magnetization is  independent of the number of neighbours. Differently, a model based on atomistic spins correctly accounts for the differences in the spin correlations for the different lattice structures. In the one hand, for the same values of the exchange parameters, due to the higher number of exchange-coupled spins, the critical temperature, $T_c$, will also be higher. For example, in a simple ferromagnet with $z$ n.n. and exchange coupling $J$, one finds that in the MFA, $k_BT_c=zJ/3$, while in ASD simulations
$\varepsilon k_BT_c=zJ/3$, where $\varepsilon<1$ and accounts for  the spin fluctuations neglected in the MFA\cite{GaraninPRB1996}.
For ferrimagnets, the same arguments apply, and more n.n. leads to a higher $T_c$ and the shape of the temperature dependent equilibrium magnetization is sensitive to the lattice dimension and structure and to the form of spin interactions\cite{PastukhMagnetism2023}. For this reason, we rely on atomistic spin dynamics simulations to conduct more accurate comparison between different lattice structures. 
\begin{table}[t]
\caption{Exchange parameters used in the atomistic spin dynamics model.\label{tab:exch}%
}
\begin{ruledtabular}
\begin{tabular}{cccc}
\textrm{$z$}&
\textrm{$J_{\rm{Fe-Fe}}$ [meV]}&
\textrm{$J_{\rm{Gd-Gd}}$} [meV]&
\textrm{$J_{\rm{Fe-Gd}}$} [meV]\\
\colrule
6 & 21.18 & 8.30  & -7.38\\
8 & 15.50 &  6.06 & -5.39\\
20 & 5.22 & 1.97  & -1.80 \\
\end{tabular}
\end{ruledtabular}
\end{table}

\begin{figure}[!t]
\includegraphics[width=8.2cm]{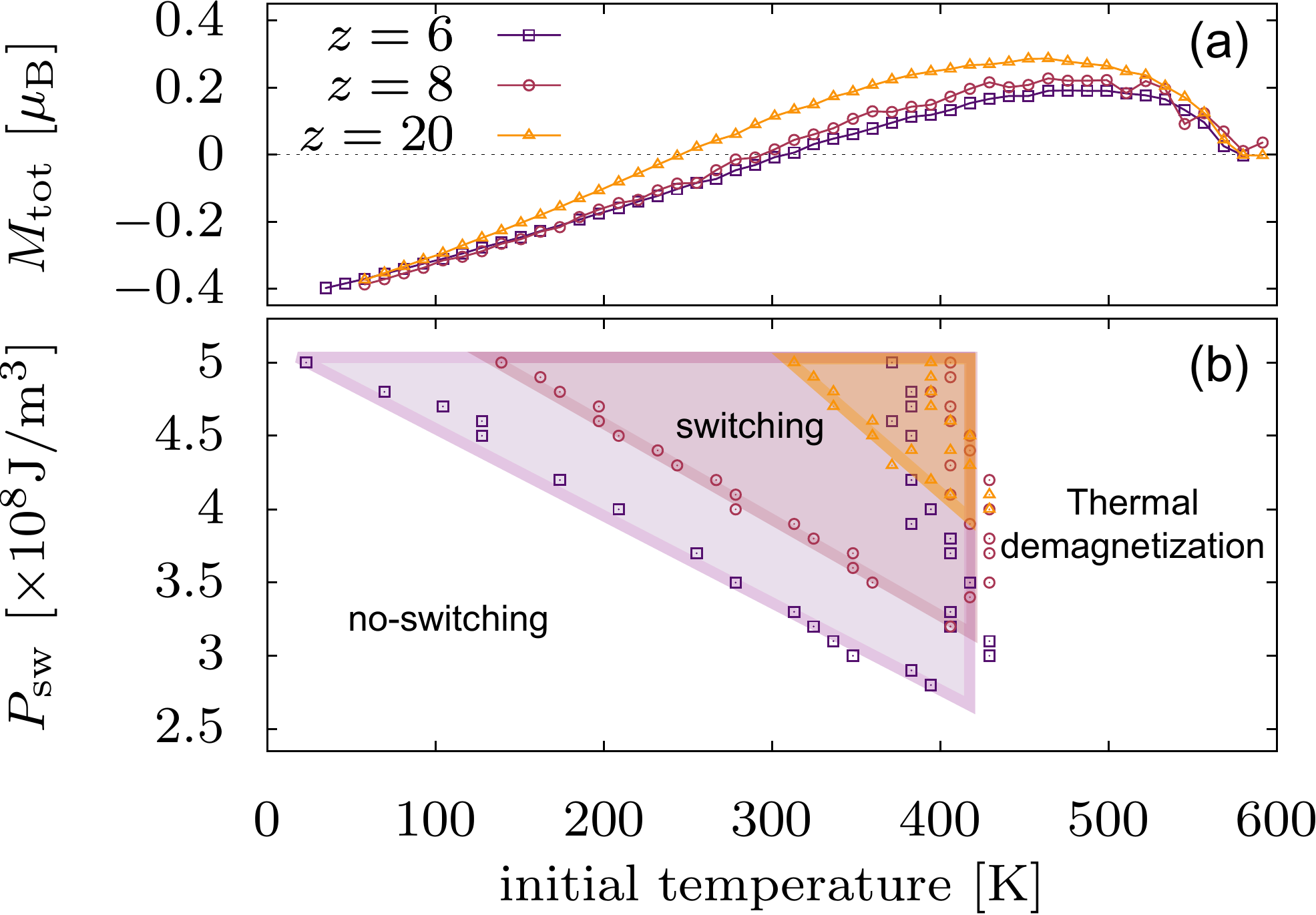}
\caption{(a) Temperature-dependent equilibrium magnetization, $M_{\rm{tot}}=M_{\rm{Gd}}-M_{\rm{Fe}}$ for three different number of exchange-coupled nearest neighbours, $z=6,8$ and 20. (b) Colored area corresponds to the values of the laser energy $P_{\rm{sw}}$ for switching. For temperature above around 400 K, only thermal demagnetization is observed. Below 400 K, for laser energy less than the minimum, no-switching happens. }
\label{im:2}
\end{figure}

We consider a classical Heisenberg spin Hamiltonian\cite{Nowak2007BOOK,OstlerPRB2011}:
\begin{equation}   
\mathcal{H}= - \sum_{i \neq j \\
\langle ij \rangle} J_{ij} \mathbf{s}_i \cdot \mathbf{s}_j - \sum_{i} d^z_i (s^z_i)^2.
\label{eq:Ham}
\end{equation}
Here, $|\mathbf{s}_i|=1$ and it represent the normalized classical spin vector at site $i$. The  two sublattices have different and antiparallel atomic magnetic  moments $\mu_a$ and $\mu_b$. The exchange coupling parameters have the same physical meaning as in the MFA [see Eq. \eqref{eq:MFA-LLB}].
The second term in Eq. \eqref{eq:Ham} describes the on-site uniaxial anisotropy, $d^z_i=0.5$ meV. 
We consider that the species of each atomic spin are randomly located in the lattice structure. The  dynamics of each atomic spin follows the stochastic Landau-Lifshitz-Gilbert (LLG) equation~\cite{Nowak2007BOOK}
\begin{equation}
\frac{\partial \mathbf{s}_i}{\partial t} = - \frac{|\gamma|}{(1+\lambda_i^2)\mu_{i}}\left[\left( \mathbf{s}_i \times \mathbf{H}_i \right) - \lambda_i  \left( \mathbf{s}_i \times \left(\mathbf{s}_i \times \mathbf{H}_i \right) \right)\right].
\label{eq:llg}
\end{equation}
$\lambda_i$ is the local intrinsic atomic damping and it coincides with those in Eq. \eqref{eq:alpha-a}. The effective field $\mathbf{H}_i= \boldsymbol{\zeta}_i - \frac{\partial \mathcal{H}}{\partial \mathbf{s}_i}$, where  thermal fluctuations are represented by the stochastic field $\boldsymbol{\zeta}_i$. Specifically, we use for the element-specific atomic magnetic moments, $\mu_{\rm{Fe}}=1.92\mu_{\rm{B}}$ and $\mu_{\rm{Gd}}=7.63\mu_{\rm{B}}$ ($\mu_{\rm{B}}$ is the Bohr magneton) and intrinsic damping parameters, $\lambda_{\rm{Fe}}=\lambda_{\rm{Gd}}=0.01$.  In Table \ref{tab:exch} the values of the exchange  coupling parameters for each lattice structure are given. We have re-scaled the exchange coupling parameters to obtain the same value for $T_c$ for all three structures. This is convenient since it allows to directly compare the energy efficiency of the switching process for  all three structures as $k_{\rm{B}}T_c \sim J_0$ sets the exchange energy scaling  on magnetic systems. 
Figure \ref{im:2}(a) shows the temperature-dependent total equilibrium magnetization $M_{\rm{tot}}=M_{\rm{Gd}}-M_{\rm{Fe}}$, where $M_a=q \mu_a m_a$ ($q_a$ concentration of element $a$) for the three cases studied here. In the three cases  the so-called magnetization compensation temperature $T_{M}$, the temperature at which $M_{\rm{tot}}=0$, slightly depends on the lattice structure, in comparison to the MFA for which all three cases behave the same.  
We use ASD computer simulations to investigate the minimum laser energy $P_{\rm{sw}}$ necessary to switch the magnetic state in GdFeCo. Since the critical role of $T_M$ in the switching behaviour of GdFeCo has been discussed in the literature before, we determine $P_{\rm{sw}}$ as a function of the initial temperature. We find three scenarios depending on the final magnetization of the sublattices, (i) switching is achieved when the magnetization of the sublattices have changed sign and are larger than 0.1 in length 30 picoseconds after the laser pulse is gone, (ii) no-switching, the sign of the sublattice magnetization remains the same and larger than 0.1, (iii) thermal demagnetization, otherwise, namely, the magnetization remains lower than 0.1 for large time scales. The colored area in Fig. \ref{im:2}(b) represents the laser energy that induces switching. For a particular system with $z$ exchange-coupled neighbours, the minimum energy necessary to switch decreases almost linearly as the initial temperature increases.  
For $z=8$, $P_{\rm{sw}}(z=8)$ is larger than $P_{\rm{sw}}(z=6)$ but proportional to it. By looking at Eq. \eqref{eq:alpha-ex}, one sees that the exchange relaxation rate scales as $z$, that is, $\alpha_{\rm{ex}}(z_1)/\alpha_{\rm{ex}}(z_2)$.  
We find that the ration $P_{\rm{sw}}(z_1)/P_{\rm{sw}}(z_2) \sim \sqrt{z_1/z_2}$ describes well this proportionality.    For initial temperature above around 400 K, the system is close to the critical temperature $T_c$, and the final temperature after the laser pulse goes beyond $T_c$. Therefore,  the magnetic system ends up into a thermal demagnetization state. For even larger number of neighbours, $z=20$, the laser energy would be too large so that the system will end up also into a thermal demagnetization state. 

To summarize, we have demonstrated that for ferrimagnetic GdFeCo alloys, the single-pulse magnetic switching is sensitive to the lattice dimension and structure and to the form of spin interactions. We have used a semi-phenomenological theory for a detailed discussion of the dependence of the relaxation terms -- exchange and relativistic -- on the number of exchange-coupled spins. We have validated these insights by computational simulations of the switching process using an atomistic spin model for GdFeCo. 
For example, for the limiting case of an infinity number of exchange-coupled spin, we predict that switching is not possible in ferrimagnets. At the same time, by reducing the number of exchange-coupled spins, the laser energy necessary to switch the magnetic state of GdFeCo reduces significantly.

The authors have no conflicts to disclose.

U. A. gratefully acknowledges support by grant PID2021-122980OB-C55 and the grant RYC-2020-030605-I funded by MCIN/AEI/10.13039/501100011033 and by "ERDF A way of making Europe" and "ESF Investing in your future".

The data that support the findings of this study are available from the corresponding authors upon reasonable request.

\bibliographystyle{apsrev4-1}
\bibliography{bib2}

\end{document}